\documentstyle[aps,prl,epsfig,twocolumn]{revtex}
\newcommand{\beeq}{\begin{equation}}
\newcommand{\eneq}{\end{equation}}
\newcommand{\beeqa}{\begin{eqnarray}}
\newcommand{\eneqa}{\end{eqnarray}}
\newcommand{\la}{\langle}
\newcommand{\ra}{\rangle}
\begin{document}
\twocolumn[\hsize\textwidth\columnwidth\hsize\csname
  @twocolumnfalse\endcsname
%\preprint{IMSc 97/07/31}
%\preprint{hep-th@ftp/0406170}
\title{Purification of Single Qubits and Reconstruction from Post - Measurement  State}
\author{Chirag Dhara \dag} 
\address{St. Xavier's College, Mumbai-400 001, INDIA}
\author{N.D. Hari Dass \ddag } 
\address{The Institute of Mathematical Sciences, Chennai - 600 113, INDIA}
\maketitle
\begin{abstract}
Purification of mixed states in Quantum Mechanics, by which we mean the
transformation into pure states, has been viewed as an {\it Operation} in
the sense of Kraus et al and explicit {\it Kraus Operators} \cite{kra1,kra2,kra3}  have been 
constructed for two seperate purification protocols. The first one, initially due to 
Schrodinger \cite{sch} and subsequently elaborated by Sudarshan 
et al \cite{sudar}, is based on the {\it preservation of probabilities}.
We have constructed a second protocol here based on {\it optimization of 
fidelities}. Both purification protocols have been implemented on a 
single qubit in an attempt to improve the fidelity of the purified  
post measurement state of the qubit with 
the initial pure state. We have considered both {\it complete} and 
{\it partial} measurements and have established bounds and inequalities 
for various fidelities. We show that our purification protocol leads to 
better state reconstruction, most explicitly so, when partial measurements are made.
\end{abstract}
\vskip 2pc] % end \twocolumn[...]
%\vskip 2pc] 
%\pacs{10.,11.,11.10.-z,11.30.Rd,12.15.-y,12.38.Qk,12.38.-t,12.39.Fe,
%12.40.-y,12.60.-i,12.60.Cn}
%%%%%%%%%%%%%%%% Latex File %%%%%%%%%%%%%%%%%
\section{Kraus Formalism} 
Kraus {\it et al} have given an extensive formalism to study all possible 
changes of quantum states (a general quantum state can be described by a 
density matrix that can be {\it pure} or {\it mixed}). The most remarkable 
features of this formalism are an intrinsic algebraic structure from the 
mathematical point of view as well as the physically striking result that 
any quantum state can be changed to any other quantum state through the 
so-called {\it Operations}.
An {\it Operation} $O$ is defined as follows: Consider a quantum system in 
the state $\rho_{sys}$ with a Hilbert Space ${\cal H}$ which is coupled to 
another quantum system, often called the {\it environment}, in the state 
$\rho_{E}$ 
and which has a state space ${\cal H}_{E}$. The system and the 
environment 
interact through a Unitary Evolution $U$ which acts on the total Hilbert Space 
${\cal H} \otimes {\cal H}_E$. Now some property of the environment is 
{\it selectively} measured by a projection operator $Q_E$ 
so that the state becomes:
\beeq
\hat \rho=({\bf I} \otimes Q_E) U (\rho_{sys} \otimes \rho_E)U^\dagger ({\bf I} \otimes Q_E)  
\eneq
The system is then considered as an isolated system described by the 
{\it reduced density matrix},
\beeq
\hat \rho_{sys}=Tr_{E}\hat \rho
\eneq
where the trace is taken over all possible states of the environment.
The resulting state change ~~~~~~~~~~~~~~~~~~~~~~~~~
$O:\rho_{sys} \longrightarrow \hat \rho_{sys}$ 
is called an Operation.
According to the Kraus formalism, this operation can be represented 
in terms of \emph{ Kraus operators} $A_{k}$ acting on the state space 
of the system such that
\beeq
\label{kreln}
\hat \rho_{sys}=\sum_{k \epsilon K} A_{k} \rho_{sys} A_{k}^\dagger
\eneq
As the measurement $Q_E$ is {\it selective}, the $A_k$ operators satisfy 
the {\it trace non-increasing} condition
\beeq
\sum_{k \epsilon K} A_{k} A_{k}^\dagger \leq {\bf I}
\eneq
where $K$ is some indexing set.
The operators $A_{k}$ are defined by % give appropriate reference-- look at pg 321 of Kraus' paper
\beeq
\label{defn}
(f, A_{k} g)=((f \otimes f_{k}^{E}), U (g \otimes g^{E}))
\eneq
where $f$, $g$ arbitrary vectors in the state space of the system, 
$\{f_{k}^{E} | k \epsilon K \}$ form an orthonormal basis of 
$Q_{E}{\cal H}_E$ and $g^{E}$ is the pure state in which the environment 
can be assumed to have started in.

As remarked earlier, operations can connect any given pair of density 
matrices $\{\rho_{1}, \rho_{2}\}$.
In particular, one can go from an initial mixed state to a pure state. 
This may sound counter-intuitive as the process of going from an initially 
pure state to a mixed state, as happens in quantum measurement, is seen 
as an {\it irreversible} step with an increase of (von Neumann)entropy. 
In this sense, the reverse process of going from mixed state to a pure 
state may seem impossible. But, as is clear from the Kraus formalism, 
this cannot happen in any {\it isolated system} but needs coupling to 
another system. In classical thermodynamics also, the entropy of a sub-system 
can always decrease without any violation of the second law.

The process of going from a mixed state to a pure state is called 
{\it Purification} and was allegedly first considered by Schrodinger 
\cite{sch}. There is a vast literature on this topic \cite{purerefs}. In the current literature 'Purification' is often understood to
be the process of associating a suitable {\it pure} state of a {\em larger}
system whose reduced density matrix is the mixed state one started with,
but for us purification of a mixed state is any protocol that produces
a pure state from it.

\section{Kraus Operators for Qubits}
Though only a selective measurement and one Unitary transformation was 
considered in arriving at equation (\ref{kreln}), it is straightforward 
to generalize to any type of measurement and any Unitary transformation 
in different combinations. From now onwards, we shall relax the condition 
of selectivity in measurements and consider all possible outcomes for 
measurements( we only consider {\it projective} measurements here). We 
shall also be restricting ourselves to 2-level systems 
(qubits) only. Then one needs two Kraus operators for a general operation.

Any pair of operators,
\beeq
\label{koperators}
A_0= (\alpha |0\rangle+ \beta |1\rangle)\langle0| ~~~ 
A_1= (\alpha |0\rangle+ \beta |1\rangle)\langle1| ~~~
\eneq
satisfy
\beeq 
A_{0}^\dagger A_0 + A_{1}^\dagger A_{1} = {\bf I}
\eneq
where $\alpha$ and $\beta$ satisfy $|\alpha|^2 + |\beta|^2 = 1$.
For any arbitrary density matrix $\rho_{in}$ these operators produce
\beeq
\rho_{out}=A_{0} \rho A_{0}^\dagger + A_{1} \rho A_{1}^\dagger 
=\left( \begin{array}{cc}
|\alpha|^2 & \alpha \beta^\star \\
\alpha^\star \beta & |\beta|^2
\end{array} \right) 
\eneq
Clearly, $\rho_{out}$ is a pure state and it 
is {\it independent} of the initial state $\rho_{in}$.
By eqn(\ref{defn}) the Kraus operators here are of the form,
\beeq
A_0= \langle0_E| U |0_E\rangle~~~
A_1= \langle1_E| U |0_E\rangle
\eneq
where the environment is asumed to start in the pure state $|0_E \ra$.
It is straight forward to check that the  unitary operator $U$ that generates 
the Kraus operators (\ref{koperators}) for this operation is:
\begin{eqnarray}
\label{Unitary}
U &=& ((\alpha |0\rangle + \beta |1\rangle) \langle0|)  \otimes \ |0_E\rangle \langle0_E|  \nonumber\\
& & +  ((\alpha |0\rangle + \beta |1\rangle) \langle1|)  \otimes \ |1_E\rangle \langle0_E|  \nonumber\\
& & +  (\alpha^\star |1\rangle\langle0|\ - \beta^\star|0\rangle \langle0|)   \otimes \ |0_E\rangle \langle1_E| \nonumber\\ 
& & + (\alpha^\star |1\rangle \langle1|\  -  \beta^\star |0\rangle \langle1|)  \otimes \ |1_E\rangle \langle1_E|
\end{eqnarray}

\section{Purification Protocol - A} % give appropriate reference.
 Consider some density matrix which is a mixture of two  
orthogonal states 
\beeq
\rho^{'} = p_1 \rho_1 + p_2 \rho_2 
\eneq
where, $\rho_{1}^2 = \rho_1, \rho_{2}^2 = \rho_2, tr(\rho_1 \rho_2)=0, tr\rho_1=tr\rho_2 = 1$.

The Purification Protocol discussed here is based on the principle of 
{\it preservation of probabilities}. In \cite{sudar} this was taken to mean
that the {\it overlap} of $\rho^{'}$ with $\rho_{1,2}$ is $p_{1,2}$. 
This was first discussed by Schrodinger \cite{sch}
and later elaborated by Sudarshan et al. 
Then this Purification Protocol leads to the {\em family} of pure states:
\beeq
\rho =  p_1 \rho_1 + p_2 \rho_2 + \sqrt{p_1 p_2} \frac{\rho_1 \Pi \rho_2 +\rho_2 \Pi \rho_1 }{\sqrt{tr(\rho_1 \Pi) tr(\rho_2 \Pi)}}
\eneq
where $\Pi$ is a projection which is not orthogonal to either $\rho_1$ or $\rho_2$.

If $\rho_1 = |0\ra\la 0|,\rho_2 = |1\ra\la 1|$ and $\Pi$ is of the form 
\beeq
\Pi = (\mu |0\rangle + \nu |1\rangle)(\langle0| \mu^\star + \langle1| \nu^*)
\eneq
($\mu$, $\nu$ $\neq 0$ since $\Pi$ is not orthogonal to either $|0\ra\la 0|$ or 
$|1\ra\la 1|$) then the purified state is given by
\beeq
\rho^A =  
p_1 |0\rangle \langle0| + p_2 |1\rangle \langle1| 
+ \sqrt{p_1 p_2}(e^{i \phi}|0\rangle \langle1| + h.c)
\eneq
where $\phi$ is the phase of $\mu \nu^{\star}$ and h.c stands for Hermitian 
conjugate. The reason that only this 
phase appears in the purified $\rho$ is that preservation of probabilities 
leaves only a phase left unspecified in a pure state. Different choices of 
$\phi$ lead to different purified states. There is no principle that selects 
a particular value of $\phi$.
The Kraus operators which generate this operation are of the form (\ref{koperators}):
\beeqa
A_0 &=& \sqrt{p_1}e^{i\phi} |0\rangle \langle0| + \sqrt{p_2} |1\ra \la 0|\nonumber\\
A_1 &=& \sqrt{p_1}e^{i\phi} |0\rangle \langle1| + \sqrt{p_2} |1\ra \la1|
\eneqa
The Unitary transformation $U$ generating these operators is of the 
form (\ref{Unitary}) with the substitutions:
$\alpha =  \sqrt{p_1}e^{i\phi} $  and
$\beta =  \sqrt{p_2} $. 

\section{Purification Protocol - B}
As shown in Sec. 2, any state (Pure or Mixed),
\beeq
\label{raw}
\rho =  \left( \begin{array}{clcr}
a & p \\
p^\star & 1-a
\end{array} \right) 
\eneq
can be purified to
\beeq
\rho_{out} = O(\rho) =  \left( \begin{array}{clcr}
|\alpha|^2 & \alpha \beta^\star \\
\alpha^\star \beta & |\beta|^2
\end{array} \right) 
\eneq
by using the Kraus operators of eqn(\ref{koperators}).
At this stage, $\rho_{out}$ can be any state of the system and the 
purification scheme is too general. Now we adopt a principle 
different from the one followed in Protocol- A to be able to fix 
the purified state. The Principle we adopt is that {\it the purified 
state must have maximal overlap with the mixed state we started with}. 
If we use the formula $tr (\rho_{1} - \rho_{2})^2$ for the distance 
between any two states $(\rho_{1}, \rho_{2})$, our principle is also 
equivalent to demanding that the purified state be {\it as close as possible} 
to the mixed state. The overlap between $\rho$ and $\rho_{out}$ is
${\cal F} = tr (\rho. \rho_{out})$.
Thus,
\beeq
{\cal F} = a |\alpha|^2 + p\alpha^\star \beta + p^\star \alpha \beta^\star 
+ (1-a) |\beta|^2
\eneq

Letting $|\alpha|^2 = \tilde{p}, |\beta|^2 = 1 - \tilde{p}, \alpha 
\beta^\star = \sqrt{\tilde{p} (1 - \tilde{p})} e^{-i \theta}$
and $p = |p| e^{i \phi}$ one has
\beeq
{\cal F} 
=  a \tilde{p} + (1-a) (1- \tilde{p}) 
+ 2 |p| \sqrt{\tilde{p} (1-\tilde{p})} \cos (\theta + \phi)
\eneq

The condition on the phase $\theta$ for maximizing $F$ is
$\theta = - \phi$.
Then,
\beeq
{\cal F}(\tilde{p}) =   a \tilde{p} + (1-a) (1- \tilde{p}) + 2 |p| \sqrt{\tilde{p} (1-\tilde{p})}
\eneq
which must be maximised w.r.t. $\tilde{p}$:
\beeq
\label{AA}
{\cal F}^{'}(\tilde{p}) = 2a - 1 + \frac {|p|(1 - 2 \tilde{p})}{\sqrt {\tilde{p} (1 - \tilde{p})}} = 0
\eneq
It can be shown that 
$F_{max}$ occurs for 
\beeq
\label{negative}
\tilde{p} = (1/2)(1 - \frac{(1 - 2a)}{(4 |p|^2 + (1 - 2a)^2)^{1/2}})
\eneq
The case $p = 0$ needs to be handeled seperately. In that case
\beeq
{\cal F} = a \tilde{p} + (1 - a)(1 - \tilde{p})
\eneq
If $a > 1/2$ then $\tilde{p} = 1$ gives ${\cal F}= a > 1/2$ and
$\rho_{out} =|0\ra\la 0|$.  
However, if  $a < 1/2$ then $\tilde{p} = 0$ gives ${\cal F} =1- a > 1/2$ and
$\rho_{out} = |1\ra\la 1|$.
Finally if $a = 1/2$, then $F = 1/2$ for both the above $\rho_{out}$ 's and no unique pure state can be picked. 

\section{Measurement of spin and Improvement in fidelity of post-measurement 
state with initial state by Purification}
An ensemble ($N$ copies-$N$ very large) of Spin - 1/2 particles is made.This 
ensemble is divided into 3 equal sub-ensembles and 
measurements are made of 
$S_{z}$, $S_{y}$ and $S_{x}$ on the respective sub - ensembles. 
Let $p_{1}$ be the probability for the outcome  
$|+\ra_{z}$, $p_{2}$ for 
$|+\ra_{y}$ and $p_{3}$ for $|+\ra_{x}$.

Let the basis for the 2-Dimensional Hilbert space be 
the eigen-vectors of $\sigma_{z}$,  $|+\ra_{z}$ $\equiv$ $(1,0)$ and 
$|-\ra_{z}$ $\equiv$ $(0,1)$.
In this basis 
\beeq
|\pm\ra_{x} \equiv (1/\sqrt{2}) \left( \begin{array}{c}
1 \\ \pm 1
\end{array} \right); 
|\pm\ra_{y} \equiv (1/\sqrt{2}) \left( \begin{array}{c}
1 \\\pm i
\end{array} \right)  
\eneq
Now, the three post-measurement density matrices are respectively:
\begin{eqnarray}
\rho_{1} &=& p_{1}|+\ra_{zz}\la+| + (1 - p_{1}) |-\ra_{zz}\la-| \nonumber\\
\rho_{2} &=& p_{2}|+\ra_{yy}\la+| + (1 - p_{2}) |-\ra_{yy}\la-| \nonumber\\
\rho_{3} &=& p_{3}|+\ra_{xx}\la+| + (1 - p_{3}) |-\ra_{xx}\la-| \nonumber\\
\end{eqnarray}

Now one takes an equal weightage of the three post-measurement density
matrices to give
$\rho_{msmt} = (1/3) (\rho_{1}+\rho_{2}+\rho_{3})$. 
Therefore,
\beeq
\rho_{msmt} = {1\over 6}\left( \begin{array}{cc}
{2p_{1}+2} & {(2p_{3}-1) + i (1-2p_{2})} \\
{(2p_{3}-1) - i (1-2p_{2})} & {4-2 p_{1}}
\end{array} \right) 
\eneq

Since this a {\it complete} measurement, the initial density matrix can be determined and is;
\beeq
\rho_{ini}= \left( \begin{array}{cc}
p_{1} & \frac{(2p_{3}-1) + i (1-2p_{2})}{2} \\
\frac{(2p_{3}-1) - i (1-2p_{2})}{2} & 1 - p_{1}
\end{array} \right) 
\eneq
Clearly, the relation between $\rho_{msmt}$ and $\rho_{ini}$ is, 
\beeq
\label{msmt}
\rho_{msmt} = (1/3) ({\bf I} + \rho_{ini})
\eneq
In a seperate publication we have established a result analogous to 
eqn(\ref{msmt}) for arbitrary systems with finite dim ${\cal H}$ \cite{ourpaper}. 
The Fidelity of $\rho_{msmt}$ with the initial state is :
\beeq
\label{fid1}
{\cal F} (\rho_{msmt},~  \rho_{ini}) = tr (\rho_{msmt} \rho_{ini}) 
= 2/3
\eneq
independent of $\rho_{ini}$.
Since, $\rho_{ini}$ is pure its  eigen-values  are $0$, $1$. Hence, 
the eigenvalues of $\rho_{msmt}$ are, from (\ref{msmt}), $1/3$ and $2/3$. 
Therefore $\rho_{msmt}$ can be written as,
\beeq
\rho_{msmt} ={2\over 3}|l\ra\la l|+{1\over 3}|s\ra\la s|
\eneq
Substituting in eqn(\ref{msmt}) and the completeness relation 
\beeq
{\bf I} = |l\ra \la l|+ |s\ra \la s|
\eneq
one finds
\beeq
\rho_{ini}=|l\ra\la l|
\eneq
Therefore, we have established that the eigenvector corresponding to the 
largest eigenvalue is the initial state.
The Purification of
\beeq
\rho_{msmt} = (2/3) |l\ra \la l| + (1/3) |s\ra \la s| 
\eneq
by Protocol - A is 
\beeqa
\rho_{msmt}^{(A)} &=&
 (2/3)  |l\ra \la l| + (1/3)  |s\ra \la s|\nonumber\\ 
 &+&  \sqrt{2}/3 (e^{i \phi}   |l\ra \la s|
 + e^{- i \phi} |s\ra \la\ l|)
\eneqa
Now,the fidelity is,
\beeq
{\cal F}(\rho_{msmt}^{(A)}, \rho_{ini}) = 
\la l|  \rho_{msmt}^{(A)}|l \ra 
= 2/3
\eneq
Comparing this result with (\ref{fid1}) the Purification protocol- A does not improve the fidelity of the purified post-measurement state with the initial state for complete orthogonal measurements. 

However, one could have taken a larger interpretation of protocol-A which
would preserve all the $p_1,p_2,p_3$ in which case the purified state has
to coincide with the initial state leading to a fidelity of unity.

However, protocol- B offers a different insight into the problem of complete and partial measurements. In fact, it is shown here that, for complete measurements, the initial or the pre-measurement state is the ``closest'' pure state (by the purification protocol- B) to the resultant post-measurement state (by taking an equal weightage of the three post-measurement states obtained from the three measurements).
 
Now, the purification of $\rho_{msmt}$ by protocol - B is 
\beeq
\rho_{msmt}^{(B)}= \left( \begin{array}{cc}
\tilde{p} & \sqrt{\tilde{p} (1- \tilde{p})} e^{i \phi} \\
\sqrt{\tilde{p} (1- \tilde{p})} e^{-i \phi} & 1 - \tilde{p}
\end{array} \right)   
\eneq
Here,
$a= (p_{1}+1)/3$ and $p=\frac{(2p_{3}-1) + i (1-2p_{2})}{6}$ 
Therefore, $|p|=\frac{\sqrt{(2p_{3}-1)^2 + (1-2p_{2})^2}}{6}$, $\cos(\phi)=\frac{(2p_{3}-1)/6}{\frac{\sqrt{(2p_{3}-1)^2 +  (1-2p_{2})^2}}{6}}$, $\sin(\phi)= \frac{(1-2p_{2})/6}{\frac{\sqrt{(2p_{3}-1)^2 +  (1-2p_{2})^2}}{6}}$
For complete measurements, $p_{1}$, $p_{2}$ and $p_{3}$ are related by,
\beeq
\label{reln}
(2p_{1}-1)^2 + (2p_{2}-1)^2 + (2p_{3}-1)^2 = 1
\eneq
It is easy to verify after some algebra that
\beeq
\Rightarrow \tilde{p} = p_{1};~~~ 
\sqrt{\tilde{p}(1-\tilde{p})}e^{i \phi} =  \frac{(2p_{3}-1) + i (1-2p_{2})}{2}
\eneq
In other words, $\rho_{msmt}^{(B)} = \rho_{ini}$.
\emph{The initial state is the closest pure state by Purification Protocol - B ,in fidelity, to the mixed state $\rho_{msmt}$}
This also means that protocol - B has purified the post-measurement state 
to maximum fidelity with the initial state. 

\subsection{Partial Measurements}
The real issue is in the context of partial measurements where
the initial state {\it can not} be unambiguously reconstructed.
In this section we establish the following two results: (i) the
purified state under protocol-B {\em always} has a greater fidelity
with the pre-measurement state than does the post-measurement state, (ii)
the fidelity of the purified state under protocol-B with the initial
state is {\em always} greater than that of the purified state under
protocol-A (in an unbiased average sense as protocol-A does not {\em favour}
any single pure state) except in some singular cases where the 
fidelities are the same. Thus protocol-B is the better when trying
to reconstruct the initial state from the post-measurement
state.

Suppose two measurements are made. Let $p_{1}$ and $p_{2}$ $\equiv$ measurement results $|+\ra_{z}$ and $|+\ra_{y}$ respectively.
Then post-measurement state, 
\beeq
\label{part}
\phi_{msmt} = \left( \begin{array}{cc}
(2p_{1}+1)/4 & i(1-2p_{2})/4 \\
-i(1-2p_{2})/4 & (3-2p_{1})/4
\end{array} \right)
\eneq
Let, the initial state be $\psi = \alpha|+\ra_{z} + \beta|-\ra_{z}$
Therfore, $p_{1} = |\alpha|^2$, $p_{2} = | \la\psi|+ \ra_y |^2$ which 
is equivalent to $\beta\alpha^{*}-\alpha\beta^{*}=i(2p_2-1)$.
From these relations we can compute the fidelity :
\beeq
\label{F1}
{\cal F}_{1}(\phi_{msmt}, \psi) = \frac{1}{4}((2p_{1} - 1)^2 + (2p_{2} - 1)^2 + 2)
\eneq
To purify the state by protocol - A, we can adopt the following procedure: We know that the initial state density matrix $\rho_{ini} = | \psi \ra \la \psi|$ is of the form, %give Neilsen Chuang reference
$$\rho_{ini} = I/2 + \la S_{x}\ra \sigma_{x} + \la S_{y}\ra \sigma_{y} + \la S_{z}\ra \sigma_{z}$$
and the relation is,
$$\la S_{x}\ra^2  + \la S_{x} \ra^2  + \la S_{x}\ra^2  = 1/4$$
Therefore,
\begin{eqnarray}
\la S_{x} \ra_{\pm} &=& \pm (\frac{1}{4} - \frac{(2p_{2} - 1)^2}{4} - \frac{(2p_{1} - 1)^2}{4})^\frac{1}{2} \nonumber\\
&=& \pm \frac{1}{2} [\sqrt{1 - (2p_{2} - 1)^2 - (2p_{1} - 1)^2}] \nonumber\\
& &
\end{eqnarray}
where we use the relations $\la S_{z} \ra$ $=$ $(2p_{1} - 1)/2$ and $\la S_{y} \ra$ $=$ $(2p_{2} - 1)/2$
One of the signs for $\la S_{x}\ra$ gives the initial state so that the 
fidelity is $1$ whereas the other choice gives some other state with a 
different fidelity. For argument's sake we assume that the choice of positive 
root gives the initial state. Then,
\beeq
{\cal F}_{2, a}(\rho_{ini}, \phi_{msmt, \la S_{x}\ra_{+}}^{(A)} ) = 1
\eneq
The fidelity when the negative root for $\la S_{x}\ra$ is chosen is,
\begin{eqnarray}
{\cal F}_{2, b} (\rho_{ini}, \phi_{msmt, \la S_{x}\ra_{-}}^{(A)}) &=& \frac{1}{2} + 2 \la S_{x}\ra_{-} \la S_{x}\ra_{+} \nonumber\\
& & + 2 \la S_{y}\ra^2 + 2 \la S_{z}\ra^2 \nonumber\\
&=& 1 - 4 \la S_{x} \ra^{2}
\end{eqnarray}
The average fidelity is,
\beeq
\label{Z}
{\cal F}_{2, av} = 1 - 2 \la S_{x} \ra^2
\eneq
Now, purification by protocol-B gives,
\beeq
\phi_{msmt}^{(B)}= \left( \begin{array}{cc}
\tilde{p} & e^{i\phi}\sqrt{\tilde{p} (1- \tilde{p})}  \\
e^{-i\phi}\sqrt{\tilde{p} (1- \tilde{p})} & 1 - \tilde{p}
\end{array} \right)   
\eneq
By comparing eqns. ($\ref{part}$) and ($\ref{raw}$) we get,
$a = (2p_{1} + 1)/4$ \ and \ 
$p = i(1-2p_{2})/4 \ \Rightarrow |p| = |(1 - 2p_{2})/4|$  and $e^{i \phi} = \pm i$ depending on the sign of $A_2$.
Introducing the notation $A_1=2p_1-1,A_2=2p_2-1$, we have
\beeq
(1-2 a) = -A_1/2;~~4|p|^2 + (1 - 2a)^2 = (1/4) (A_1^2+A_2^2)
\eneq
Therefore using (\ref{negative}),
\beeq
\tilde{p} = (1- \frac{A_{1}}{\sqrt{A_1^2 + A_2^2}})
\eneq
Using the relations for $|\alpha|^2$, $p_{1}$ and $p_{2}$
\begin{eqnarray}
{\cal F}_{3}(\phi_{msmt}^{(B)}, \psi) &=& \la\psi|\phi_{msmt}^{(B)}|\psi\ra \nonumber\\
&=& \frac{1}{2} (1  \nonumber\\
&+&[A_1^2 + A_2^2]^{\frac{1}{2}}) \nonumber\\
& & 
\end{eqnarray}
Then the Fidelity ${\cal F}_{3}$ is:
\beeq
{\cal F}_{3} = \frac{1}{2} [1 + (A_{1}^2 + A_{2}^2)^{1/2}]
\eneq
Therefore,
\begin{eqnarray}
{\cal F}_{3} - {\cal F}_{1} &=& \frac{1}{4} [2(A_{1}^2 + A_{2}^2)^{1/2} - (A_{1}^2 + A_{2}^2)] \nonumber\\
& \geq& 0 
\end{eqnarray}
In other words,
$${\cal F}_{3} \geq {\cal F}_{1}$$
This means that for a partial measurement where only two components of spin are measured, the fidelity can always be improved over ${\cal F}_{1}(\phi_{msmt}, \rho_{ini})$ by Purification Protocol - B.
Furthermore,
$${\cal F}_{3}(\phi_{msmt}^{(B)}, \psi) = \frac{1}{2} [1 + (1 - 4 \la S_{x} \ra^2)^{1/2}]$$
Now from $(\ref{Z})$,
$${\cal F}_{2, av} = \frac{1}{2} (1 + 1 - 4 \la S_{x} \ra^2))$$
Hence,
$$2 {\cal F}_{2, av} - 1 = 1 - 4 \la S_{x} \ra^2$$
and this leads to
$$2 {\cal F}_{3} - 1 = (2 {\cal F}_{2, av} - 1)^{1/2}$$
Clearly  since  $ (2 {\cal F}_{2, av} - 1)$ $\leq 1$ we have,
$${\cal F}_{3} \geq {\cal F}_{2, av}$$ 
Clearly, Purification by Protocol - B reconstructes the state with better fidelity than does purification by protocol - A on the {\it average}.

When only one component of spin is measured, say, $S_{z}$ then we have the p = 0 case as has been worked out in the last part of section 4. 
Here,
$$\chi_{msmt} = p_{1} |+\ra_{zz} \la+| + (1 - p_{1})|-\ra_{zz} \la-|$$
Again, the initial state is of the form: $\psi = \alpha |+\ra_{z} + \beta |-\ra_{z}$. Now, we know only that $|\alpha|^2 = p_{1}$. 
The Fidelity of the post-measurement state with the initial state is:
\begin{eqnarray}
{\cal F}_{4}(\chi_{msmt}, |\psi\ra \la \psi|) &=& \la\psi|\chi_{msmt}|\psi\ra \nonumber\\
&=& p_{1}^2 + (1 - p_{1})^2 \nonumber\\
\end{eqnarray}

Now, if we purify the $\chi_{msmt}$ by protocol - A, then
\beeq
\chi_{msmt}^{(A)} = \left( \begin{array}{cc}
p_{1} & \sqrt{p_{1}(1 - p_{1})}e^{i \phi} \\
\sqrt{p_{1}(1 - p_{1})}e^{-i \phi} & 1 - p_{1}
\end{array} \right)
\eneq
The $\rho_{ini} = |\psi\ra \la \psi|$ is of the form:
\beeq
\rho_{ini} = \left( \begin{array}{cc}
p_{1} & \sqrt{p_{1}(1 - p_{1})}e^{-i \theta} \\
\sqrt{p_{1}(1 - p_{1})}e^{i \theta} & 1 - p_{1}
\end{array} \right)
\eneq
where, $p_{1}$ is known from measurement, but the phase $\theta$ cannot be determined. All values of $\theta$ should be considered equally likely. Therefore,
\begin{eqnarray}
{\cal F}_{5}(\chi_{msmt}^{(A)}, \rho_{ini}) &=& tr(\chi_{msmt}^{(A)} \rho_{ini}) \nonumber\\
&=& p_{1}^2 + (1 - p_{1})^2 \nonumber\\
& & + 2p_{1} (1 - p_{1}) \cos(\theta + \phi)   
\end{eqnarray}
The average Fidelity with equal weightage for all $\theta$ is,  
$${\cal F}_{5, av} = p_{1}^2 + (1 - p_{1})^2$$ 

Now, if $p_{1}$ $\geq$ $1/2$, then the purified state according to protocol - B is,
$$\chi_{msmt}^{(B)} = \left( \begin{array}{cc}
1 & 0 \\
0 & 0
\end{array} \right)$$
Then,
\begin{eqnarray}
{\cal F}_{6}(\chi_{msmt}^{(B)}, |\psi\ra \la \psi |) &=& tr(\chi_{msmt}^{(B)}  |\psi\ra \la\psi| ) \nonumber\\
&=& p_{1}
\end{eqnarray}
Since, $p_{1}$ $\geq$ $1/2$, it can be verified that 
$${\cal F}_{6}(\chi_{msmt}^{(B)}, | \psi\ra \la \psi|) \geq {\cal F}_{4}(\chi_{msmt}, |\psi\ra \la \psi|)$$
It can be verified that even for $p_{1}$ $<$ $1/2$ protocol - B always leads to an improvement in fidelity.

Also, since ${\cal F}_{5, av}$ $=$ ${\cal F}_{4}$, we have the relation
$${\cal F}_{6}(\chi_{msmt}^{(B)}, |\psi\ra \la \psi |) \geq {\cal F}_{5, av}
(\chi_{msmt}^{(A)}, \rho_{ini})$$
\emph{Thus, the fidelity offered by protocol - B is better than the average fidelity offered by protocol - A.}

\section{Acknowledgements}
CD thanks Prof. Ajay Patwardhan of St. Xavier's College, Mumbai, 
for his invaluable encouragement, support and guidance over the years as also 
for his commitment towards his students. CD also thanks The Institute of 
Mathematical Sciences for its hospitality and support through a Fellowship 
under the Visiting Students Programme.

%\end{references}
\end{document}